\documentclass[]{jfm}

\usepackage{graphicx}
\usepackage[justification=justified]{caption}

\usepackage{newtxtext}
\usepackage{newtxmath}
\usepackage{natbib}
\usepackage{hyperref}
\usepackage[noabbrev,nameinlink,capitalise]{cleveref}
\hypersetup{
    colorlinks=true,
    urlcolor=blue,
    citecolor=blue,
    linkcolor=blue,
}

\usepackage{subfigure}

\usepackage{siunitx}

\usepackage{orcidlink}

\newcommand{\RomanNumeralCaps}[1]
\linenumbers

\title{The drag of a filament moving in a supported spherical bilayer}

\author{Wenzheng Shi\aff{1},
  Moslem Moradi\aff{1}
 \and Ehssan Nazockdast \aff{1}\corresp{\email{ehssan@email.unc.edu}}}

\affiliation{\aff{1}Department of Applied Physical Sciences, The University of North Carolina at Chapel Hill, Chapel Hill, North Carolina 27599, USA}

\begin{document}
\maketitle
\begin{abstract}
Many of the cell membrane’s vital functions are achieved by the self-organization of the proteins and biopolymers embedded in it. The protein dynamics are in part determined by its drag. A large number of these proteins can polymerize to form filaments.\emph{In-vitro} studies of protein-membrane interactions often involve using rigid beads coated with lipid bilayers, as a model for the cell membrane. Motivated by this, we use slender-body theory to compute the translational and rotational resistance of a single filamentous protein embedded in the outer layer of a supported bilayer membrane and surrounded on the exterior by a Newtonian fluid. We first consider the regime, where the two layers are strongly coupled through their inter-leaflet friction. We find that the drag along the parallel direction grows linearly with the filament's length and quadratically with the length for perpendicular and the rotational drag coefficients. These findings are explained using scaling arguments and by analyzing the velocity fields around the moving filament. We, then, present and discuss the qualitative differences between the drag of a filament moving in a freely suspended bilayer and a supported membrane as a function of the membrane's inter-leaflet friction. Finally, we briefly discuss how these findings can be used in experiments to determine membrane rheology. In summary, we present a formulation that allows computing the effects of membrane properties (its curvature, viscosity, and inter-leaflet friction), and the exterior and interior 3D fluids' depth and viscosity on the drag of a rod-like/filamentous protein, all in a unified theoretical framework.
\end{abstract}



\section{\label{sec:Intro}Introduction}
\begin{figure}
\centerline{\includegraphics[width=0.7\textwidth]{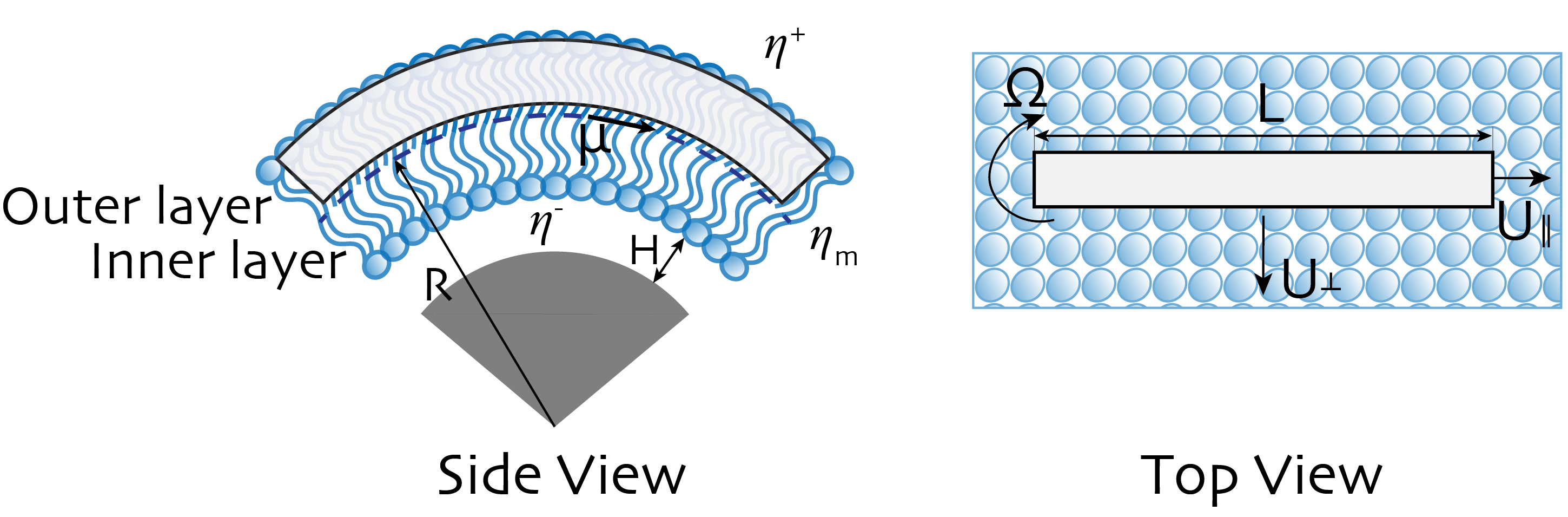}}
\centering
\hspace{0cm}
\caption{A schematic representation of the problem studied here. A filament is embedded in the outer layer of an incompressible spherical bilayer membrane of viscosity $\eta_{\mathrm{m}}$ and surrounded on the exterior side and interior side by 3D Newtonian fluid of shear viscosity $\eta^{+}$ and $\eta^{-}$, respectively. The inner leaflet is solid sphere of radius $R$. The substrate is separated from the adjacent leaflet with by a thin nanoscopic layer of fluid of depth $H$. $\mu$ is the friction coefficient between the leaflets. The filament dynamics is described by three modes of motions: translation along its axis ($U_{\parallel}$), translation perpendicular to its axis ($U_{\perp}$), and rotation around its center ($U_{\Omega}$). 
\label{fig:illustration}}
\end{figure}

The transport of proteins and biopolymers in biological membranes is an important step in determining their organization \citep{Alberts2022molecular}. Membrane proteins can pass across the bilayer thickness (transmembrane proteins), or interact with one of the leaflets (monotopic proteins). Monotopic proteins can polymerize to form filaments and other higher-order structures that span the micron-scale membrane surface \citep{ khmelinskaia2021membrane, baranova2020diffusion}. In many processes, the protein function is determined by its organization \citep{shi2023curvature}. 
Simplified \emph{in-vitro} systems are powerful tools for increasing our physical understanding of complex biological systems, including the organization of membrane proteins. Supported bilayers, rigid beads coated with lipid bilayers, are widely used as a model for spherical cell membranes \citep{cannon2019amphipathic, bridges2014septin}. The diffusion and transport of these semiflexible filamentous proteins within the fluid membrane is determined, in part, by their hydrodynamic drag. Here, we present the translational and rotational drag of a single filament moving in a spherical supported bilayer, as shown in \cref{fig:illustration}. 
\par
A starting point for analyzing protein lateral motion in biomembranes is the work of \citet{Saffman1976} (see also \citet{Saffman1975}, and \cite{hughes1981translational}), which gives an  expression for the drag coefficient of a disk of radius, $a$, moving in an infinite planar membrane of 2D viscosity $\eta_m$, and surrounded with an infinite 3D Newtonian fluid of shear viscosity $\eta_f$ on both sides. 
The coupling between the membrane and 3D fluid domains introduces Saffman-Delbr\"uck (SD) length $\ell_0=\eta_m/\eta_f$, which is the length over which momentum transfers from 2D membrane to 3D bulk fluids. 
For small particles ($a/\ell_0\ll1$), \citet{Saffman1976} showed that the drag coefficient is only a weak logarithmic function of the disk radius: $\xi_\text{Saff}=4\pi\eta_m\left(\ln (2\ell_0/a)-\gamma\right)^{-1}$ where $\gamma$ is Euler–Mascheroni constant. Safmman's results, and simple extensions of it, have been used to measure membrane rheology in microrheological experiments; see \cite{molaei2021interfacial}, \cite{kim2011interfacial}, \cite{prasad2006two} and  Chapter 4 of \cite{morozov2015introduction}. 
\par
\citet{Evans1988} (see also \citet{sackmann1996supported}) extended Saffman's work to a disk moving in a planar membrane that is supported on a rigid boundary. The effect of this boundary is modeled using a Brinkman-like friction term, $\mu \mathbf{u}_m$ in the membrane momentum equation, where $\mu$ is the friction coefficient and $\mathbf{u}_m$ is the membrane tangential velocity. The friction introduces a new length scale: ${b}=\sqrt{\eta_m/\mu}$. \citet{Stone1998} considered the case of a planar membrane overlaying a 3D fluid domain of finite depth, $H$, which similarly introduces a length scale defined as $\ell_H=\sqrt{\ell_0 H}$. 
In both cases, the drag coefficient asymptotes to Saffman's results for small particles ($a/b\ll 1$ or $a/\ell_{H}\ll 1$), with $b$ or $\ell_H$ replacing $\ell_0$ in expression for the drag coefficients. Furthermore, in the likely scenario of $b/a\ll 1$ or $\ell_H/a\ll 1$, both models predict a quadratic increase in drag with respect to the particle size ($\xi \propto \eta_m (a/b)^2$ or $\xi \propto \eta_m (a/\ell_H)^2$). 
\citet{stone2015mobility} used reciprocal theorem and perturbation analysis to compute the drag on a spherical or oblate spheroidal particle moving in the membrane and protruding into the subphase fluid. 
\citet{zhou2022drag} computed the drag on a sphere in a similar setup, where the particle is trapped at the interface of two fluids where they considered the effects of the gravity and interfacial deformation. 
\par
\citet{Levine2004} used a slender-body theory to compute the translational and rotational drag of a rod-like inclusion moving in a planar membrane and adjacent to infinite bulk fluids. They found that when $L/\ell_0\gg 1$, the drag in all directions scales with the bulk fluid viscosity, and linearly with the filament length with an extra weak logarithmic dependency in parallel direction: $\xi_{\perp,\Omega}\propto \eta_f L $ and $\xi_\parallel \propto \eta_f L/\ln(L/\ell_0)$. These predictions were found to be in good agreement with the experiments in the range $0.01 \le L/\ell_0 \le 10$ \citep{lee2010combined, Klopp2017}. 
\citet{fischer2004drag} generalized the work of \citet{Levine2004}, to a planar membrane overlying a fluid domain of finite depth. They found that when $H/\ell_{0}\ll1$, the parallel drag grows linearly with $L/\ell_{H}$ while the perpendicular drag grows superlinearly. 
\par
Most theoretical studies, including the ones surveyed thus far, consider inclusions that fill the entire membrane thickness.
We know that monotopic proteins typically bind to one of the two leaflets in lipid bilayers. Motivated by this observation, \citet{Camley2013} computed the drag of a disk embedded in the top leaflet of a planar membrane and surrounded by the infinite 3D bulk fluid on the outer side and finite bulk fluid on the interior. The two leaflets are coupled through a friction term. They found that the drag monotonically increases with the inter-leaflet friction coefficient with results matching those of \cite{Evans1988}, when inter-leaflet friction is replaced with the substrate's friction. 

\par
While the majority of theoretical studies on the hydrodynamic drag of inclusion in membranes have focused on planar membranes, in most biological applications membranes take a spherical or more complex curved geometry. 
\cite{Henle2010} considered the drag of a disk moving in a spherical membrane that is surrounded by bulk fluids on both sides. 
They found that the drag follows the results of \citet{Saffman1975}, as long as SD length is replaced with $\min(\ell_{0},R)$ where $R$ is the radius of the membrane;
see also \cite{manikantan2020tunable}, \cite{samanta2021vortex}, and \cite{jain2023force} for studies on the surface flow and aggregations induced force and torque dipoles.  
\par
In a recent study, we used slender-body theory to compute the drag of a filament bound to a spherical monolayer immersed in 3D bulk fluids on the interior and exterior \citep{shi2022hydrodynamics}. Our computations show that the closed spherical geometry gives rise to flow confinement effects that increase in strength with increasing the ratio of the filament's length to membrane radius $L/R$. 
These effects only cause mild increases in the filament's parallel and rotational resistance; hence, the resistance in these directions can be quantitatively mapped to the results on a planar membrane when the momentum transfer length-scale is modified to  $\ell^\star=(\ell_0^{-1}+R^{-1})^{-1}$. 
In contrast, we find that the flow confinement effects result in a superlinear increase in perpendicular drag with the filament's length when $L/R >1$. These effects are absent in free space planar membranes.  

This study extends our previous work to a filament embedded in the outer leaflet of a  bilayer membrane that is supported by a rigid sphere on the interior, as shown in \cref{fig:illustration}. 
We present the conservation equations in \S \ref{sec:Formula} and present the closed-form fundamental solution to a point force in this general geometry in \cref{sec:appA}. We use these solutions in a slender-body theory to compute the translational and rotational resistance of a filament in \S \ref{sec:results}. Finally,  summarize and discuss our main findings in \S \ref{sec:summary}.    

\section{Formulation} \label{sec:Formula}
We consider a filament of length $L$, embedded in the outer leaflet of a lipid bilayer that is supported on the interior by a rigid sphere of radius $R$. Both leaflets have 2D shear viscosity of $\eta_m$. The bilayer is surrounded by a semi-infinite 3D fluid of viscosity $\eta^+$ on the exterior. We assume the rigid boundary is separated from the lipid head groups of the bottom leaflet by a thin nanoscopic layer of fluid of viscosity $\eta^{-}$ \citep{sackmann1996supported} and depth $H$, as shown in \cref{fig:illustration}.
The two leaflets are coupled through a friction body force that is proportional to the relative velocity of the two leaflets. We assume the filament curvature is constant along its length and equal to $1/R$, \footnote{This is the most likely conformation of the filament if the intrinsic curvature of the filament is smaller than the sphere and the bending forces are much larger than thermal and inter-particle forces.}, which decouples the translational and rotational motion of the filament, due to geometric and flow symmetries. Thus, the translational resistance tensor is defined as $\boldsymbol{\xi}=\xi_\parallel \mathbf{q}\mathbf{q}+ \xi_\perp\left(\mathbf{I}-\mathbf{q}\mathbf{q}\right)$, where $\mathbf{I}$ is the identity matrix and $\mathbf{q}$ is the filament's unit alignment vector. 
The rotational resistance, $\xi_{\Omega}$, is independent of $\mathbf{q}$.
\par
Assuming flow incompressibility on the membrane and 3D fluid domains and negligible inertia, the associated momentum and continuity equations for the membrane and 3D fluid domains are \citep{samanta2021vortex, Henle2010,shi2022hydrodynamics}: 

\begin{subequations}
\begin{align}
\label{eq:stokes_3D}
&\eta^{\pm}\nabla^{2}\mathbf{u}^{\pm}-\nabla p^{\pm}=\mathbf{0}, &
&\nabla\cdot\mathbf{u}^{\pm}=0,& \\
\label{eq:stokes_outer}
&\eta_{m}\left(\Delta_\gamma \mathbf{u}_{m}^{\text{o}}
+K(\mathbf{x}_m)\mathbf{u}_{m}^{\text{o}}-\frac{1}{b^{2}}(\mathbf{u}_{m}^{\text{o}}-\mathbf{u}_{m}^{\text{i}})\right)-\nabla_\gamma p_{m}^{\text{o}} +\mathbf{T}^{\text{o}}
=\mathbf{0},& 
&\nabla_\gamma\cdot \mathbf{u}_{m}^{\text{o}}=0, \\ 
\label{eq:stokes_inner}
&\eta_{m}\left(\Delta_\gamma \mathbf{u}_{m}^{\text{i}}
+K(\mathbf{x}_m)\mathbf{u}_{m}^{\text{i}}-\frac{1}{b^{2}}(\mathbf{u}_{m}^{\text{i}}-\mathbf{u}_{m}^{\text{o}})\right)-\nabla_\gamma p_{m}^{\text{i}} +\mathbf{T}^{\text{i}}
=\mathbf{0},& &\nabla_\gamma\cdot \mathbf{u}_{m}^{\text{i}}=0, 
\end{align}
\label{eq:Eqs}
\end{subequations}
where $\mathbf{u}^{\pm}$ and $p^{\pm}$ are the velocity and pressure fields in 3D fluid domains, and $\mathbf{u}_{m}^{\text{o}}$, $\mathbf{u}_{m}^{\text{i}}$ and $p_{m}^{\text{o}}$, $p_{m}^{\text{i}}$ are the velocity and pressure fields in the outer-layer and inner-layer of the membrane, respectively; $\Delta_\gamma$ and $\nabla_{\gamma}\cdot$ are the surface (defined by $\gamma$) Laplacian and Divergence operators, $K$ is the local Gaussian curvature of the surface, $b=\sqrt{\eta_{m}/\mu}$, where $\mu$ is the inter-leaflet drag coefficient, $\mathbf{T}^{\text{o}}=\boldsymbol{\sigma}^{+}(\mathbf{x}_m)|_{r=R}\cdot \mathbf{n}(\mathbf{x}_m)$, and $\mathbf{T}^{\text{i}}=-\boldsymbol{\sigma}^{-}(\mathbf{x}_m)|_{r=R}\cdot \mathbf{n}(\mathbf{x}_m)$ are the traction applied from the surrounding 3D fluid domains on the membrane from exterior and interior flow, respectively, where $\boldsymbol{\sigma}^\pm$ denotes the 3D fluid stress and $\mathbf{n}(\mathbf{x}_m)$ is the surface normal vector pointing towards the exterior domain. 
\par
The boundary conditions (BCs) are the continuity of velocity and stress fields across all interfaces. The stress continuity is automatically satisfied by adding the traction terms from 3D fluids to membrane momentum equations. The velocity and stress fields decay to zero at infinitely large distances from the interface in the outer fluid domain, $\lim\limits_{r\to \infty} u_{\theta,\phi}^+ (r)\to 0$. Finally, the velocity at the boundary of the supported solid sphere is zero $\mathbf{u}_{m}^{\text{in}}|_{r=R-H}=\mathbf{0}$. Since we take the interior to be rigid, the radial velocity becomes exactly zero across all layers and 3D fluid domains. 
\par
The constant Gaussian curvature on the sphere, $K=R^{-2}$, significantly simplifies \cref{eq:Eqs}. As a result, we can find closed-form expressions of Green's function and compute the membrane velocity fields at an arbitrary point $(\theta,\phi)$ in response to a point-force at $(\theta_0,\phi_0)$: $\mathbf{u}_m(\theta,\phi)=\mathbf{G}(\theta-\theta_0,\phi-\phi_0)\cdot \mathbf{f}(\theta_0,\phi_0)$. Here, $\mathbf{G}$ is the Green's function, and $\theta \in [0,\pi]$ and $\phi \in (0,2\pi]$ are the polar and azimuthal angles in spherical coordinates. The detailed derivation of Green's function and the final expressions are presented in \cref{sec:appA}. 
\par

It is very reasonable to assume that $H/R\ll1$ in almost all applications. In \cref{sec:appA} we show that in this regime the effects of the inner leaflet and the thin fluid layer on the top leaflet can be combined into a single effective friction lengthscale: $b^{\star}=\sqrt{\ell^{-}H+b^{2}}$, where $\ell^{-}=\eta_{m}/\eta^{-}$. 
Below we provide a simple scaling analysis that bears this result.
\par
When $H/R\ll1$, the flow inside the thin fluid layer can be approximated as simple shear flow, which results in the associated traction on the bottom leaflet to scale as 
$\mathbf{T}^\text{i}\sim \eta^{-} \mathbf{u}^\text{i} /H \sim \eta_m \mathbf{u}^\text{i}/(\ell^{-} H)$. When the fluid layer thickness is the smallest length scale, the drag from the fluid layer is of the same order of magnitude or larger than the drag force from membrane viscosity: $|\mathbf{T}^\text{i}|\ge |\eta_m \nabla^2 \mathbf{u}_m^\text{i}|$. 
Furthermore, we have $K=R^{-2}\ll (\ell^{-} H)^{-1}$. Thus, in our scaling analysis, we can drop the first two terms in \cref{eq:stokes_inner} in comparison with $\mathbf{T}^\text{i}$, and we get:
$\left(\mathbf{u}^\text{o}_m-\mathbf{u}^\text{i}_m\right)/b^2 \sim \mathbf{u}^\text{i}_m/(\ell^{-} H)$, which gives 
$$ \mathbf{u}^\text{o}_m \sim \left(1+\frac{b^2}{\ell^{-} H}\right)\mathbf{u}^\text{i}_m.$$
We now can eliminate $\mathbf{u}_m^\text{i}$ from \cref{eq:stokes_outer}, by replacing the term $b^{-2}\left(\mathbf{u}_m^\text{o}-\mathbf{u}_m^\text{i}\right)$ with $(b^\star)^{-2}\mathbf{u}_m^\text{o}$ using the above scaling. Following these steps we recover $b^\star=\sqrt{\ell^{-} H +b^2}$.   
As a result, \cref{eq:Eqs} simplify to
\begin{subequations}
\begin{align}
\label{eq:stokes_3D2}
& \eta^{+}\nabla^{2}\mathbf{u}^{+}-\nabla p^{+}=\mathbf{0}, &\nabla\cdot\mathbf{u}^{+}=0,& \\
\label{eq:stokes_outer2}
& \eta_{m}\left(\Delta_\gamma \mathbf{u}_{m}^{\text{o}}
+K\mathbf{u}_{m}^{\text{o}}-\frac{\mathbf{u}_{m}^{\text{o}}}{{b^\star}^{2}}\right)-\nabla_\gamma p_{m}^{\text{o}} +\mathbf{T}^{\text{o}}=\mathbf{0}, &\nabla_\gamma\cdot \mathbf{u}_{m}^{\text{o}}=0.
\end{align}
\label{eq:Eqs2}
\end{subequations}
The BCs are the continuity of the velocity and stress of outer 3D fluid and membrane of the outer layer. Also, 3D fluid velocity and stress decay to zero at infinitely large distances.  
\par
Since we have a closed-form solution of Green's function for \cref{eq:Eqs} and the more special case of \cref{eq:Eqs2}, we can use slender-body theory to model the flow disturbances induced by a filament with a distribution of force densities. To calculate the resistance in parallel ($\xi_\parallel$), perpendicular ($\xi_\perp$), and rotational($\xi_{\Omega}$) directions, we set the filament velocity(vorticity) as a constant in each direction and compute the distribution of force density on the filament by solving the following integral equation:
\begin{equation} 
\label{eq:FiberMobility}
\mathbf{U}(s)= \int_{-L/2}^{L/2}\mathbf{G}
(\mathbf{X}(s)-\boldsymbol{X}(s^\prime))\cdot  \mathbf{f}(s^\prime)\mathrm{d}s^\prime,
\end{equation}
where $\mathbf{X}(s)$ is a point located at $s$ arclength of the filament and $\mathbf{G}(\mathbf{X}(s)-\boldsymbol{X}(s^\prime))$ is the Green’s function of the membrane-outer 3D fluid coupled system in response to a point-force applied on the membrane at position $\mathbf{X}(s^\prime)$. The details of the numerical implementations are given in our earlier work \citep{shi2022hydrodynamics}.
Integrating the force densities (torque densities) along the filament's length gives the total force (torque), which is equal to the drag in each direction for a unit translational (rotational) velocity. 
We assume that the filament thickness, $a$, is negligible compared to all the other lengths. The error of the resistance due to the thickness of the filaments, unlike the filament in 3D flow, scales with $\mathcal{O}(\epsilon)$, where $\epsilon=a/L$, and thus here we model filament as an ideal 1D line; see error analysis in \citet{shi2022hydrodynamics}. 
\par
Applying a net force to a spherical membrane leads to a net torque on the membrane and its interior, which leads to a rigid body rotation of the spherical membrane \citep{Henle2010, samanta2021vortex}. This effect is not present in a planar membrane. 
The resistance is defined based on the \emph{relative} velocity of the filament with respect to the ambient fluid: $\mathbf{F}=\mathbf{R}\cdot (\mathbf{U}-\mathbf{u}^\infty)$, where $\mathbf{R}$ is the filament's resistance tensor and $\mathbf{u}^\infty$ is the membrane's rotational velocity due to the net torque on it; see details in \cref{sec:appA}. 
\section{Results} 
\label{sec:results}
After combining the effect of the thin fluid layer and the bottom leaflet into a single friction coefficient, the filament's drag only depends on four lengths: $L$, $R$, $\ell^+=\eta_m/\eta^+$ and $b^\star$.
We begin by considering a strong coupling between the leaflets: 
$b^\star/\min{(R,\ell^{+})}\ll1$. Recall that $b^\star>\max(\sqrt{\ell^{-}H}, b)$. 
Hence, in the strongly coupled limit, $\max{(b,\sqrt{\ell^{-}H})}$ 
are significantly smaller than $\min{(R,\ell^{+})}$. Hereafter, the $\star$ superscript is dropped for brevity.

\subsection{Two leaflets are strongly coupled} \label{sec:resist_couple}
\begin{figure}
\begin{subfigure}[] 
{
\begin{minipage}{0.3\textwidth}
\centering
\hspace{0cm}
\includegraphics[width=0.9\textwidth]{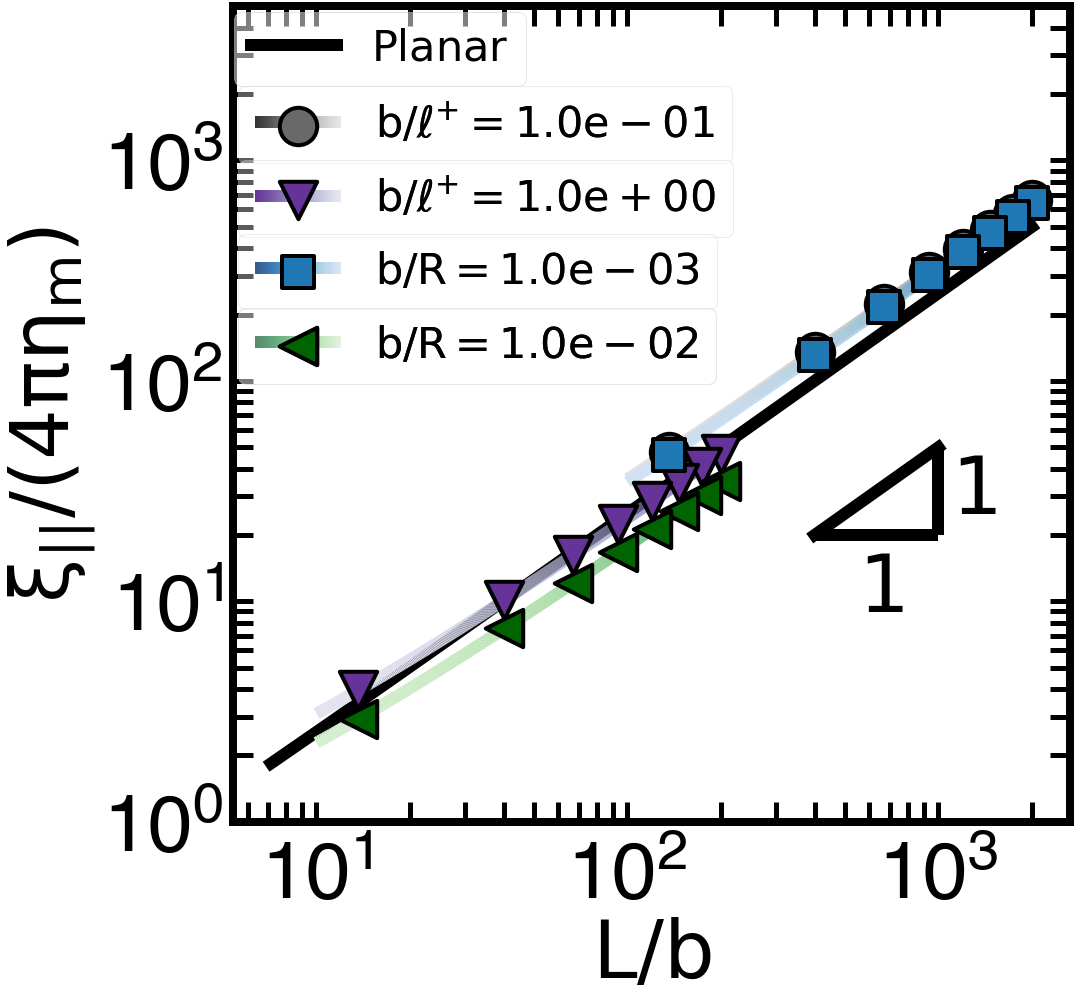}
\label{fig:pa_H0_b_lR001100}
\end{minipage}
}
\end{subfigure}
\begin{subfigure}[] 
{
\begin{minipage}{0.3\textwidth}
\centering
\hspace{-0.6cm}
\includegraphics[width=0.9\textwidth]{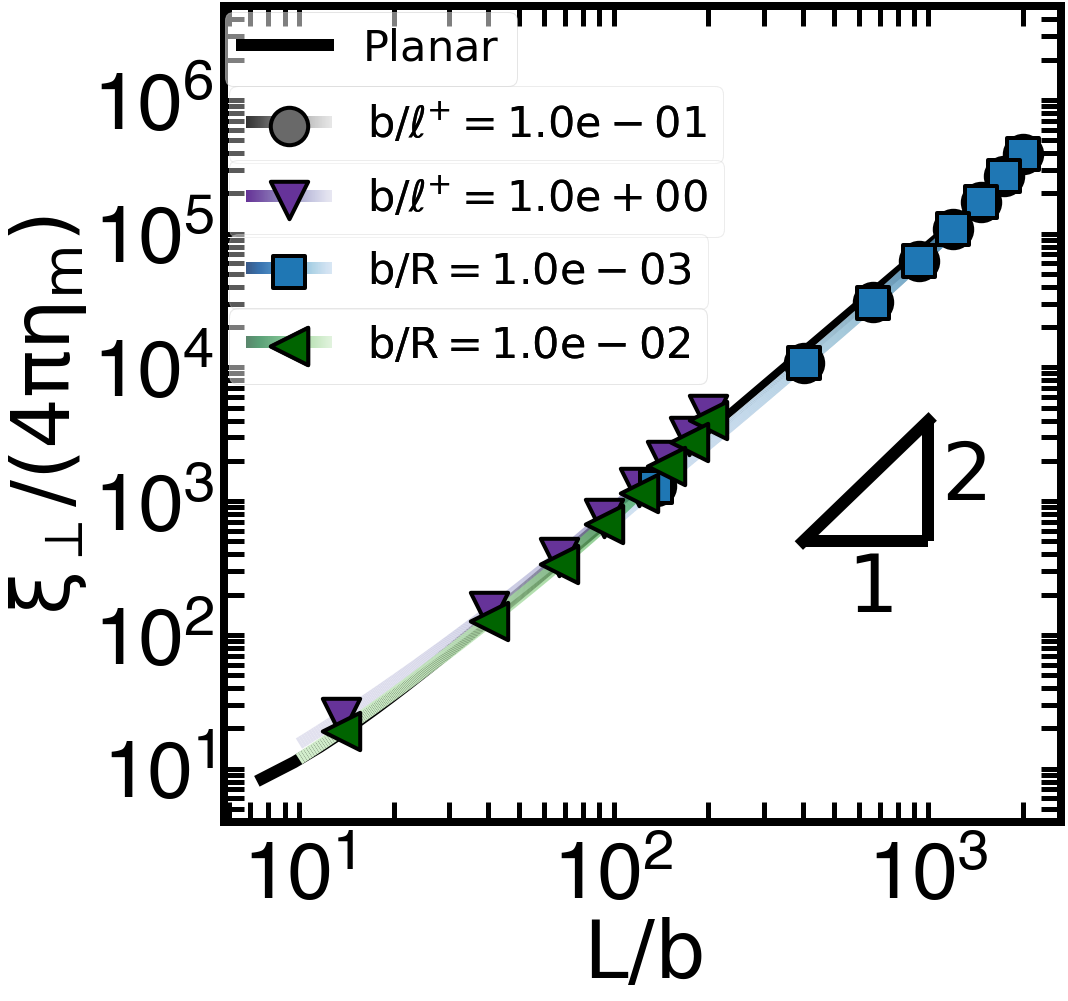}
\label{fig:pe_H0_b_lR001100}
\end{minipage}
}
\end{subfigure}
\begin{subfigure}[] 
{
\begin{minipage}{0.3\textwidth}
\centering
\hspace{-1.1cm}
\includegraphics[width=0.9\textwidth]{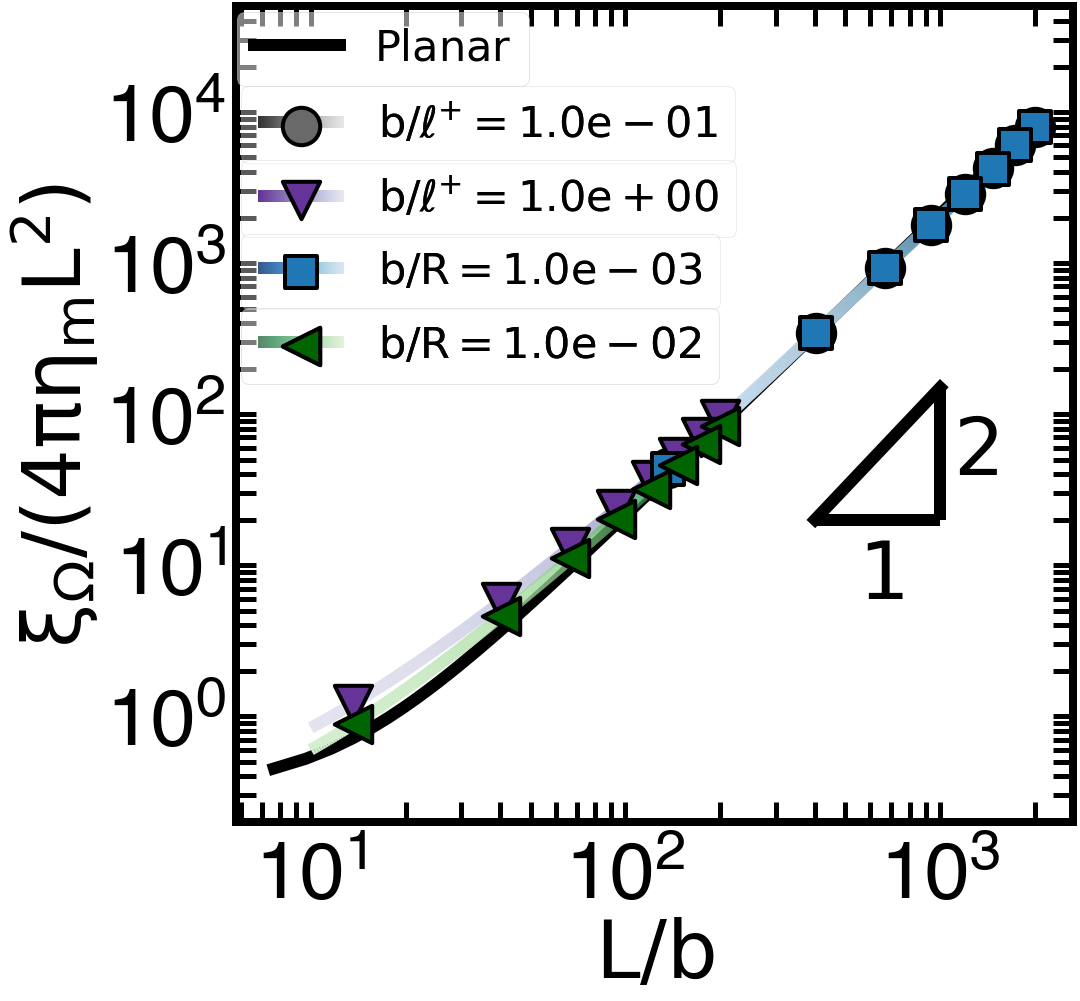}
\label{fig:ro_H0_b_lR001100}
\end{minipage}
}
\end{subfigure}
\caption{The dimensionless parallel (a), perpendicular (b), and rotational (c) drag coefficients as a function of $L/b$. 
Gray(circle) and purple(downward triangle) symbols represent different ratios of $b/\ell^{+}$, while $\ell^{+}/R=\num{1e-2}$ was kept fixed. 
Blue(square) and Green(leftward triangle) represent different ratios of $b/R$ while $\ell^{+}/R=\num{1e2}$ was kept fixed. 
The solid black lines the associated resistance values of Brinkman flow in planar membranes, where the x-axis is $L/\sqrt{\kappa}$ and $\kappa$ is the permeability of the porous medium.
\label{fig:resist_couple}}
\end{figure}

\cref{fig:resist_couple} shows the computed values of parallel, perpendicular, and rotational resistance (drag coefficients) as a function $L/b$ for different values of $b/\ell^{+}\ll 1$ and $b/R\ll 1$. We expect the resistance to be determined by the ratio of the filament's length to the shortest hydrodynamic screening length, i.e., $b$. 
Indeed, all the curves collapse on a single curve as a function $L/b$ in parallel, perpendicular, and rotational directions as long as $b\ll\min(\ell^{+}, R)$.
In this regime, all the resistance functions converge to the case of a filament embedded in a 2D planar Brinkman flow, which is plotted as a black line in \cref{fig:resist_couple}. Because $b\ll\min(\ell^{+}, R)$, the \cref{eq:stokes_outer2} can be simplified into:
\begin{equation} \label{eq:brinkman2d}
    \eta_{m}\left(\Delta_\gamma \mathbf{u}_{m}^{\text{o}}-\frac{\mathbf{u}_{m}^{\text{o}}}{b^{2}}\right)-\nabla_\gamma p_{m}^{\text{o}} =\mathbf{0}, \quad\quad \nabla_\gamma\cdot \mathbf{u}_{m}^{\text{o}}=0,
\end{equation}
where $b^{2}$ plays the same role as the permeability in porous media \citep{Brinkman1949}. The contributions from the curvature and 3D bulk flow are negligible when the resistance is dominated by the inter-leaflet friction \citep{Camley2013}. 
\par
Note that we have only presented the results for $L/b \ge 1$. For $L/b<1$ the drag is dominated by the membrane shear stresses and the drag, as expected, asymptote to the Saffman's formulae with $b$ replacing $\ell_0$. When, $L/b\gg 1$, the parallel and perpendicular resistance, exhibit a linear and quadratic dependency with $L$: $\xi_{\|}\propto{L/b}$ and  $\xi_{\perp}\propto{(L/b)^{2}}$. 

To explain this scaling, let's first assume the force distribution along the filament is nearly uniform due to its high aspect ratio. As a result, the filament's mobility scales with $\chi_{\parallel,\perp}=\xi_{\parallel, \perp}^{-1} \sim \frac{1}{L}\int_0^{L} G_{\parallel, \perp} (r)dr$.  Also, the integral of the Green's function of 2D Brinkman flow satisfies: $\lim_{\tilde{r}\to \infty} \int_{0}^{\tilde{r}} G_{\|}(r)dr = \pi-1/\tilde{r}$ and $\lim_{\tilde{r}\to \infty} \int_{0}^{\tilde{r}}G_{\perp}(r)dr=1/\tilde{r}$ \citep{kohr2008green}. Thus, when $L/b\gg1$, the filament’s mobility scales with $\xi^{-1}_{\|}\sim\frac{b}{L}\int_{0}^{L/b}G_{\|}(\Tilde{r})d\Tilde{r}=\mathcal{O}(b/L)$ and $\xi^{-1}_{\perp}\sim\frac{b}{L}\int_{0}^{L/b}G_{\perp}(\Tilde{r})d\Tilde{r}=\mathcal{O}(b/L)^{2}$, where $\Tilde{r}=r/b$. This scaling leads to the linear and quadratic growth of resistance in parallel and perpendicular directions, respectively. The rotational resistance also scales quadratically, $\xi_{\Omega}\propto{(L/b)^{2}}$, since Green's function in the perpendicular direction is used in computing it. 
\par
\begin{figure}
\centering
\hspace{0cm}
\includegraphics[width=0.85\textwidth]{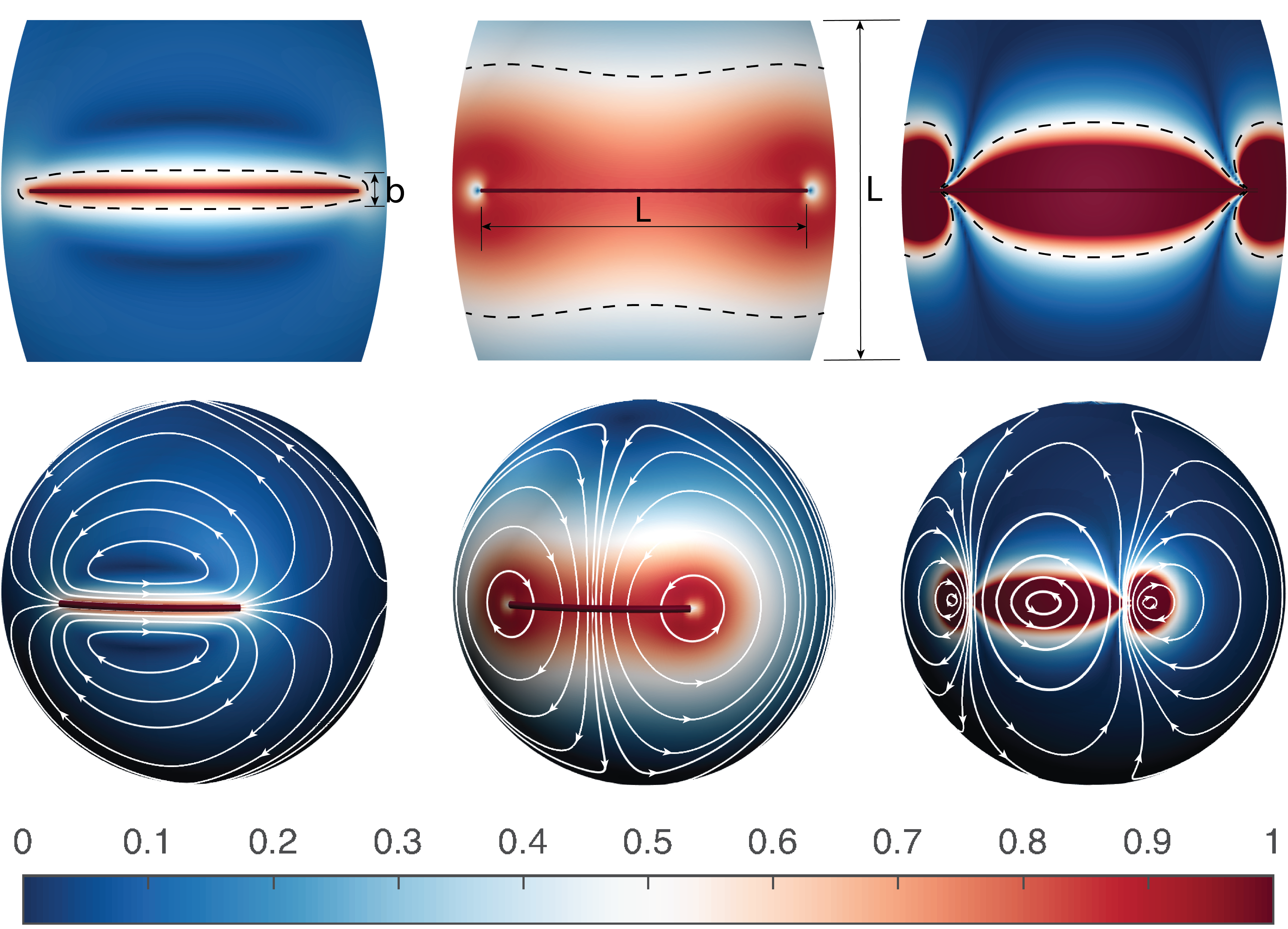}
\caption{The interfacial flow field induced by filament motion in parallel (left), perpendicular (middle), and rotational (right) directions, when $\ell^{+}/R=1$, $L/R=1$ and $b/R=0.1$. The scale for the filament's filament's translational rotational velocity is set to be 1. The white lines represent the streamlines and the underlying heatmap represents the magnitude of the fluid velocity (a-b) and vorticity (c). The upper row shows the zoomed-in velocity (left and middle) and vorticity (right) fields close to the filament, where the black dashed lines represent the contour of velocity $\lvert\mathbf{u}_{m}\rvert=0.5$ and vorticity $\lvert\mathbf{\Omega}_{m}\rvert=0.5$. 
\label{fig:flow}}
\end{figure}

To gain a better physical understanding of these scaling relationships it is useful to study the the velocity field generated by the filament motion. Figure \ref{fig:flow} shows the flow streamlines for parallel, perpendicular, and rotational motions. The colormap underlying the streamlines shows the velocity(vorticity) magnitude on the spherical surface when normalized by the net velocity(vorticity) of the filament. These results are presented for the choice of $L/R=1$, $\ell^+/R=1$, and $b/R=0.1$. As it can be seen in the left column of \cref{fig:flow}, the velocity magnitude decays to zero very rapidly around the filament moving in the parallel direction; see also the dashed contour corresponding to $|\mathbf{u}_m|=0.5$. An inspection of the flow fields shows that the velocity fields decay over distances that scale with $b$. Thus, we can approximate the system as a rectangle with $L\times b$ hydrodynamic dimensions moving with velocity $U_\parallel$. Given that the traction from membrane flow gradients scales with membrane velocity magnitude, and that these gradients are very small outside of the rectangle, we can safely ignore those contributions to the drag compared to the traction from the bottom leaflet/substrate. As a result, the total drag force on the rectangle is simply the integral of the substrate traction, $f=\eta_m U_\parallel/b^2$, over the area of the rectangle, $L\times b$. So we get $F_\parallel \approx (Lb) \eta_m U_\parallel/b^2= \eta_m U (L/b)$, which yields $\xi_\parallel=F/U_\parallel\sim \eta_m (L/b)$. 
\par
The middle row of \cref{fig:flow} shows the streamlines and the colormaps of velocity magnitude when the filament moves perpendicular to its axis. Notice that, unlike the velocity fields for parallel motion, the velocity magnitudes remain of $\mathcal{O}(1)$ over distances that scale with the length of the filament.
Hence, the effective dimensions of the filament scale as $L\times L$. Following the same line of arguments as in parallel motion, we can approximate the total drag from inter-leaflet frictional forces as $F_\perp \sim (L^2)\eta_m U_\perp/b^2$, which gives $\xi_\perp\sim \eta_m (L/b)^2$. We note that the perpendicular drag has the same form as the drag of disk of size $L$ when $L/b \gg 1$ \citep{Evans1988, sackmann1996supported}. It is easy to explain this similarity by noting that the effective hydrodynamic dimensions of a filament of length $L$ are the same as a disk of the same diameter. The same line of arguments can be used to explain why we observe the same scaling for the rotational drag as well:  $\xi_\Omega \sim \eta_m (L/b)^2$. 

\subsection{The drag coefficients in supported bilayers vs vesicles} \label{sec:resist_varyb}
The cell membrane's interior can range from a nearly solid to a fluid interface, depending on the particular cellular process and proteins involved. As such, vesicles are also extensively used as synthetic models of the cell membrane \citep{walde2010building}.  
In this section, we explore the differences in protein dynamics in vesicles and supported bilayers by comparing the drag coefficients in these two systems. The derived Green's function for \cref{eq:Eqs} applies to arbitrary values of $R,\, H$, and $\ell^\pm$ in their accepted physical range.  
Thus, we can compute the drag of a filament moving in the outer leaflet of a vesicle, $\xi^\text{vsc}$, by setting $H=R$ in the Green's function and solving \cref{eq:FiberMobility}. 
We have performed these calculations for the same values of $\ell^+$, $R$, and $b$ that are reported in \cref{fig:resist_couple}, but with $H=R$. For simplicity, we assumed the interior and exterior fluids have the same viscosity i.e. $\ell=\ell^+=\ell^-=\eta_m/\eta^{\pm}$. 
\par
In an earlier study, we computed the drag on a filament moving in a suspended lipid \emph{monolayer}, assuming the same 3D viscosity on the interior and exterior \citep{shi2022hydrodynamics}. Our computations show that the ratio of the computed drag on a vesicle to the drag on a monolayer membrane remains in the range $[0.5-2]$, over the entire range of parameters and that drag coefficients in both systems have the same scaling with the filament's length. To keep the focus on the large differences between supported vs vesicles/monolayer, we do not discuss $\mathcal{O}(1)$ variations of the drag coefficients between monolayer and vesicles in the main text, and provide the relevant results and analysis in \hyperlink{Supp}{Supplementary Materials} Figure S2. 
\par
The scaling relationships of the filament's drag embedded in a monolayer/vesicle are summarized in 
\cref{tab:vsc_mono}. When the filament's length, $L$, is smaller than 
$\ell^{\star}=\min(\ell_{0},R)$, the drag converges to  the results of \citet{Saffman1976}, with  $\ell^\star$ replacing $\ell_0$ as the shortest hydrodynamic screening length. 
When $\ell_{0}\ll{L}<R$, the drag asymptotes to the drag of a long filament, $L/\ell_0 \gg 1$, embedded in a planar membrane \citep{Levine2004}. Finally, when the filament length is larger than the sphere radius, $L>R$, the closed spherical geometry gives rise to flow confinement effects that lead to an increase in the perpendicular drag and superlinear scaling with $L/\ell^\star$; these flow confinement effects are significantly weaker in parallel and rotational directions.   

\begin{table}
  \begin{center}
\def~{\hphantom{0}}
  \begin{tabular}{lccc}
      Different asymptotic limits  & $\hat{\xi}_\parallel^{\text{vsc, mono}}$   &   $\hat{\xi}_{\perp}^{\text{vsc, mono}}$ & $\hat{\xi}_{\Omega}^{\text{vsc, mono}}$ \\[5pt]
       $L\ll\ell^{\star}$ \citep{Saffman1976}   & $\mathcal{O}\left(\frac{1}{\ln(\ell^{\star}/L)}\right)$ & $\mathcal{O}\left(\frac{1}{\ln(\ell^{\star}/L)}\right)$ &  $\mathcal{O}\left(1\right)$ \\ [8pt]
       $\ell_{0}\ll{L}<R$ \citep{Levine2004}  & $\mathcal{O}\left(\frac{L/\ell_{0}}{\ln(L/\ell_{0})}\right)$ & $\mathcal{O}\left(L/\ell_{0}\right)$ & $\mathcal{O}\left(L/\ell_{0}\right)$ \\ [8pt]
       $L>R$ \citep{shi2022hydrodynamics}  & $\mathcal{O}\left(\frac{L/\ell^{\star}}{\ln(L/\ell^{\star})}\right)$ & $\mathcal{O}\left((L/\ell^{\star})^{\alpha}\right)$ & $\mathcal{O}\left(L/\ell^{\star}\right)$ \\ 
  \end{tabular}
  \caption{The scaling behavior of  dimensionless drag coefficients of a rod-like particle of length, $L$, moving in suspended membrane monolayer and bilayers (vesicles). Here, $\ell^{\star}=\min(\ell_{0},R)$ and $1<\alpha \le 2 $;  $\hat{\xi}_{\parallel, \perp} =\xi_{\parallel, \perp}/4\pi \eta_m $ and 
  $\hat{\xi}_{\Omega} =\xi_{\Omega}/4\pi \eta_m L^2$.}
  \label{tab:vsc_mono}
  \end{center}
\end{table}

We can now study the changes in the ratio of the filament's drag in supported and suspended spherical membranes as a function of the other rations of physical lengths in the system.
We present the results for \emph{small sphere} limit, by taking $\ell^{+}/R=\num{1e2}$, but the discussions and the scaling behavior of the drag ratios hold for all values of $\ell^{+}/R=\num{1e-2}$. In this limit, the drag in all directions becomes nearly independent of $\ell^{+}$ and only a function of $L/R$ and $b/R$.  
In \hyperlink{Supp}{Supplementary Materials} Figure S1 we provide the same results for the \emph{large} sphere limit, where $\ell^{+}/R=\num{1e-2}$. The behavior and the different scaling relationships are very similar to the small sphere limit, with $\ell^{+}$ replacing $R$ in all scaling behaviors.      
\par
\begin{figure}
\begin{subfigure}[] 
{
\begin{minipage}{0.3\textwidth}
\centering
\hspace{-0.1cm}
\includegraphics[width=0.99\textwidth]{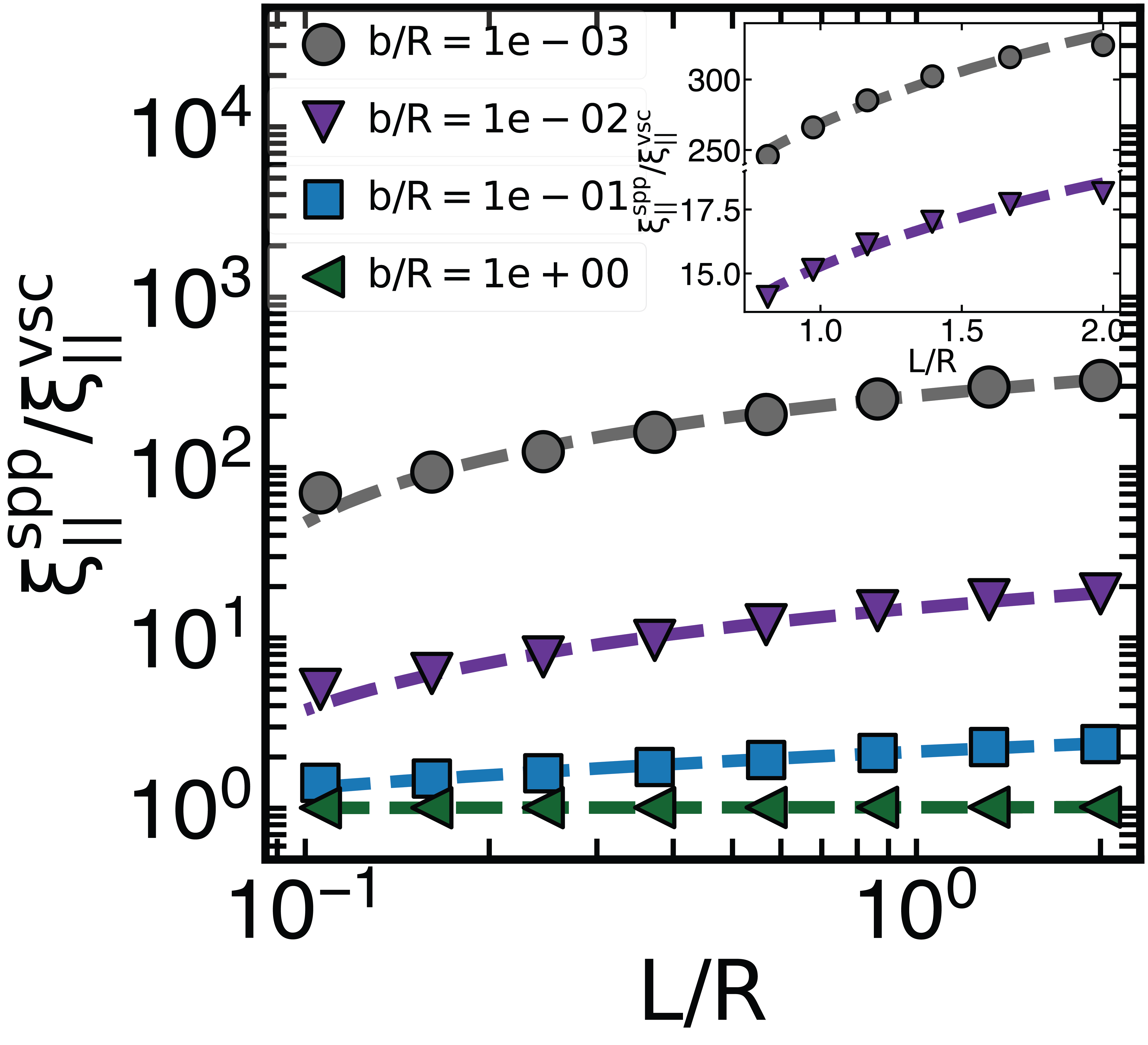}
\label{fig:pa_ratio_HR1_lR100}
\end{minipage}
}
\end{subfigure}
\begin{subfigure}[] 
{
\begin{minipage}{0.3\textwidth}
\centering
\hspace{-0.1cm}
\includegraphics[width=0.99\textwidth]{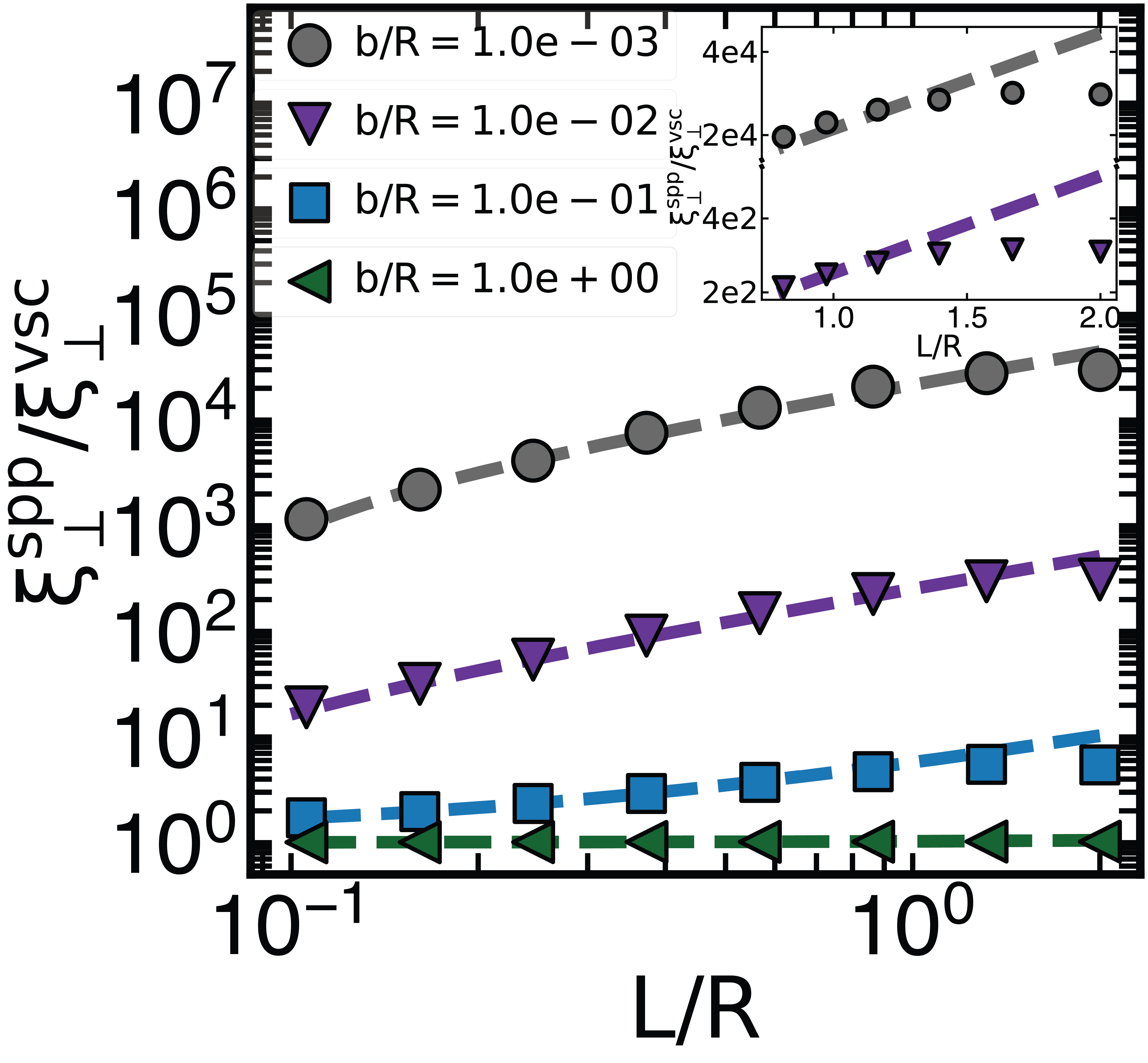}
\label{fig:pe_ratio_HR1_lR100}
\end{minipage}
}
\end{subfigure}
\begin{subfigure}[] 
{
\begin{minipage}{0.3\textwidth}
\centering
\hspace{-0.1cm}
\includegraphics[width=0.99\textwidth]{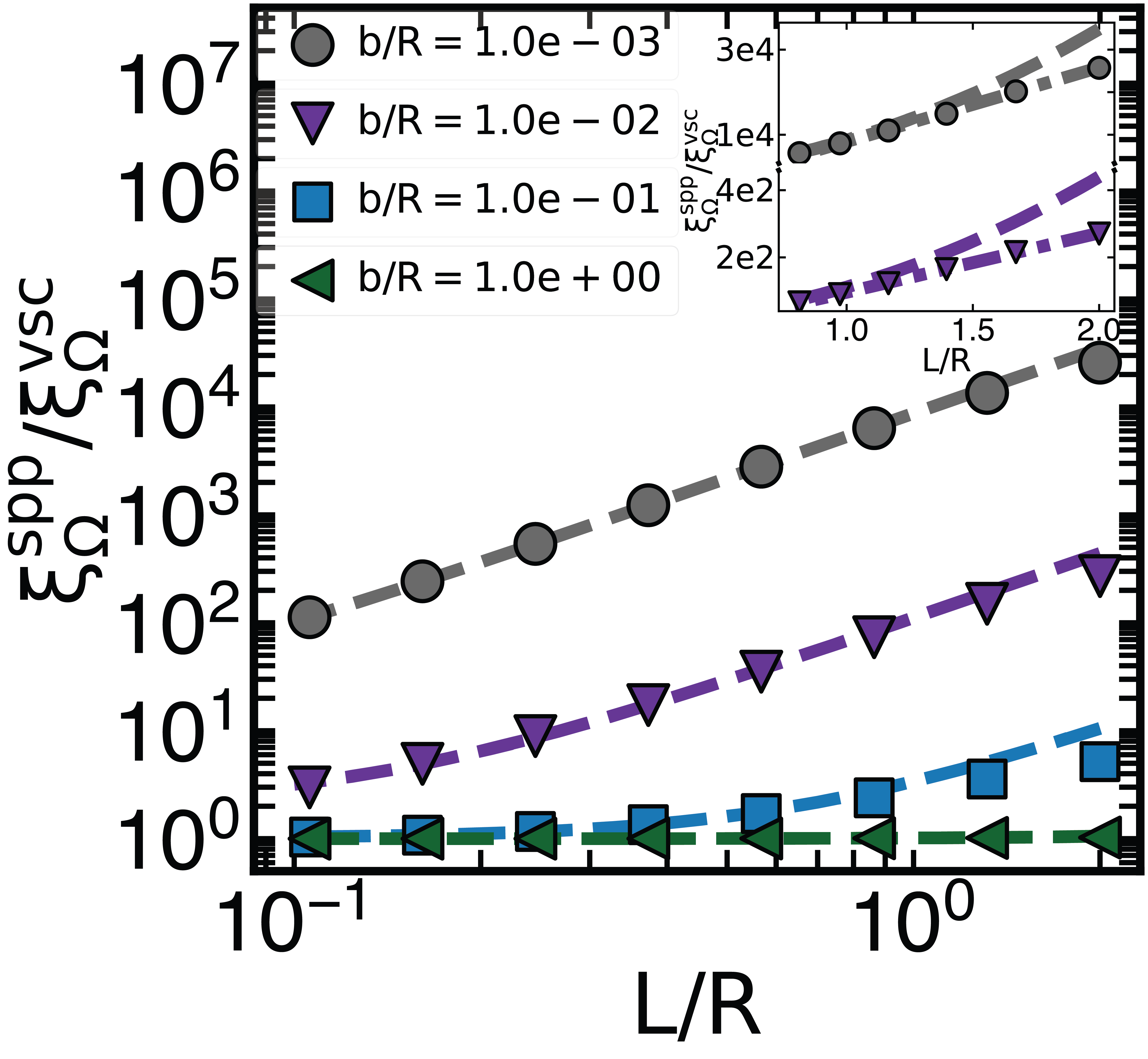}
\label{fig:ro_ratio_HR1_lR100}
\end{minipage}
}
\end{subfigure}
\caption{The ratio of the parallel (a), perpendicular (b), and rotational (c) drag of a filament moving in the outer layer of a spherical supported bilayer to the drag of the same  filament in a vesicle (freely suspended bilayer) as a function of the ratio of the filament's length to the membrane radius, $L/R$.
The results are presented for the case of \emph{small} sphere, $\ell^{+}/R=100$, where the drag becomes nearly independent of $\ell^{+}$. 
The dashed lines are the fits corresponding to  $y=a\log(x)+b$, $y=ax+b$, and  $y=ax^{2}+b$ in \cref{fig:pa_ratio_HR1_lR100,,fig:pe_ratio_HR1_lR100,,fig:ro_ratio_HR1_lR100}, respectively.
The inset figures show the deviation of the scaling laws as the filament length is increased to  $L/R>1$. The dashed lines are the same fitting functions as the main plot. The dash-dotted line in the inner plot of (c) is a linear fitting function $y=ax+b$. 
%
\label{fig:ratio_varyb}}
\end{figure}

\cref{fig:ratio_varyb} shows the ratio of drag coefficients vs $L/R$ in parallel, perpendicular, and rotational directions. The results are presented for a wide range of $b/R$ ratios. When $b/R>1$, the inter-leaflet coupling is weak, and the drag is determined only by the traction from the outer fluid in the supported bilayer and the vesicle alike. As a result, the ratios remain close to one.  
As it can be seen in \cref{fig:pa_ratio_HR1_lR100}, the ratios of parallel resistance strongly increase with decreasing $b/R$. When $b/R\ll 1$, we observe a logarithmic scaling of the ratios with $L/R$; see dashed lines in \cref{fig:pa_ratio_HR1_lR100}. 
This scaling can be explained by recalling that $\xi_{\parallel}^\text{spp} \sim L/b$ and $\xi_{\parallel}^\text{vsc}\sim (L/R)\left(\ln (L/R)\right)^{-1}$, which makes their ratio scale as $\xi_{\parallel}^\text{supp}/\xi_{\parallel}^\text{vsc} \sim (b/R)^{-1}\ln (L/R)$. 
\par
\cref{fig:pe_ratio_HR1_lR100} presents the resistance ratios vs $L/R$ in the perpendicular direction for the same values of $b/R$. Again, the drag ratios reduce to $1$ for weak couplings of the two leaflets ($b/R\gg 1$). The ratios show a strong increase with decreasing $b/R$, with these increases being stronger than the parallel case (compare the values corresponding to $b/R=\num{1e-3}$ in both cases). The ratio 
$\xi_{\perp}^\text{spp}/\xi_{\perp}^\text{vsc}$shows a linear scaling with $L/R$ when $L/R<1$, compared to the logarithmic scaling we observed for the parallel drag; see dashed lines in \cref{fig:pe_ratio_HR1_lR100}. 
This scaling can similarly be explained by noting that 
$\xi_{\perp}^\text{spp} \sim (L/b)^2$ and $\xi_{\perp}^\text{vsc}\sim (L/R)$. Thus, we have $\xi_{\parallel}^\text{spp}/\xi_{\parallel}^\text{vsc} \sim (b/R)^{-2} (L/R)$. As a result, for a fixed value of $L/R$, the 
ratio increases as $(b/R)^{-2}$ as $b$ decreases, and at a fixed value of $b/R$ the ratio scales as $L/R$. 
Note that when $L/R>1$, as shown in the inner plot of \cref{fig:pe_ratio_HR1_lR100}, the slope of the plots begins to decrease and reach a plateau at a fixed value of $b/R$. To explain this we recall that flow confinement effects lead to superlinear growth of perpendicular resistance with $L/R$ in vesicles. As a result, $\xi_{\perp}^\text{supp}/\xi_{\perp}^\text{vsc}\sim (L/R)^{2-\alpha}$ with $1<\alpha \le 2$. 
\par
\cref{fig:ro_ratio_HR1_lR100} shows the rotational resistance ratios vs $L/R$. 
As we discussed earlier, when $L/R<1$ (or more generally $L/\ell^\star<1$), $\xi_{\Omega}^\text{vsc} \sim \mathcal{O}(1)$ (see \cref{tab:vsc_mono}), while $\xi_\Omega^\text{spp} \sim (L/R)^2$ making $\xi_\Omega^\text{spp}/\xi_\Omega^\text{vsc} \sim (L/R)^2$. This trend is shown with the fitted dashed lines of the form $y=ax^2+b$ in the same figure. When $L/R>1$, $\xi_{\Omega}^\text{vsc} \sim (L/R)$, which results in a linear scaling of the ratio. This is shown as a dash-dotted line in the inner plot of \cref{fig:ro_ratio_HR1_lR100}. 
\section{Summary}
\label{sec:summary}
The transport and assembly of rod-like proteins and cytoskeletal filaments  on biological membranes occur in many cellular processes.
A widely used \emph{in-vitro} setup for studying protein-membrane interactions involves coating rigid beads of comparable size to the cell with lipid bilayers, as models for the cell membrane. Different aspects of the protein structure and dynamics can be measured over time using different microscopy techniques \citep{cannon2019amphipathic}. 
As previous studies on planar-supported membranes have shown \citep{sackmann1996supported}, the presence of a rigid substrate qualitatively changes the tangential flows and the resulting hydrodynamic drag of the inclusions. However, except for a few studies \citep{Levine2004, Henle2010}, these studies have been limited to disk-like particles and planar membranes.
This work fills some of the gaps in the literature by computing the drag of a single filament in a supported spherical bilayer, using a slender-body formulation. Furthermore, the formulation presented here combines the effects of membrane curvature, inter-leaflet friction of lipid bilayers, and the depth and viscosity of the surrounding 3D domains into a unified theoretical framework. 
\par
To summarize, our results show that the presence of the rigid substrate influences the filament's dynamics in several distinct ways. 
When the two leaflets of the bilayer are strongly coupled, the drag becomes independent of membrane radius and Saffman-Delbr\"uck  length and only a function of the inter-leaflet friction length scale, $b$. We find that the drag in the parallel direction scales linearly with the filament's length, while the perpendicular and rotational drags scale quadratically with length; see details in \cref{fig:resist_couple}. We explained this scaling using the analytical form the Green's function for a point-force on the membrane, and by visualizing the velocity fields around the filament undergoing parallel, perpendicular, and rotational motion; see \cref{fig:flow}.  
\par
We compute the ratio of the drag in supported bilayers to freely suspended vesicles, $\xi^\text{spp}/\xi^\text{vsc}$. 
When inter-leaflets are weakly coupled, the drag in all directions asymptotes to the drag on monolayer membranes and vesicles. Increasing the inter-leaflet friction leads to significant increases in the ratio of drag coefficients, particularly in perpendicular and rotational directions. 
We explained these variations in terms of the scaling between $\xi^\text{spp}$ vs $b/R$ and $\xi^\text{vsc}$ vs $L/R$ for parallel, perpendicular, and rotational drag coefficients.  
\par
The measurement of membrane viscosity has been conducted through decades by many different experimental methods and simulations. The reported values vary within the range $\eta_{m}\in[10^{-10}, 10^{-6}]\,(\mathrm{Pa}\cdot\mathrm{s}\cdot\mathrm{m})$, depending on lipid composition \citep{block2018brownian, sakuma2020viscosity, nagao2021relationship}.
In comparison, fewer studies have been conducted to quantify the inter-leaflet friction coefficients \citep{pott2002dynamics, den2007intermonolayer, jonsson2009mechanical, botan2015mixed, zgorski2019surface, amador2021hydrodynamic}. The reported values have a very wide range $\mu\in[10^{0}, 10^{9}]\,(\mathrm{Pa}\cdot\mathrm{s}\cdot\mathrm{m}^{-1})$, due to a number of experimental uncertainties and challenges (see Chapter 4 of \cite{morozov2015introduction}). 
Our results suggest another method of measuring inter-leaflet friction and membrane viscosity in the likely condition that the inter-leaflet friction length is the smallest hydrodynamic length of the system, $b\ll \min(\ell_0,R)$. In this limit, the dimensionless drag in the perpendicular direction scales as $\xi_{\perp}/(4\pi\eta_{m})\sim (L/b)^2$. Since $b^2=\eta_m/\mu$, we get $\xi_\perp \sim \mu L^2$ i.e. the perpendicular drag is independent of membrane viscosity and only a function of the filament's length and inter-leaflet friction coefficient. 
Applying the same analysis to parallel drag gives the following scaling:  $\xi_{\|}\sim L\sqrt{\mu \eta_{m}}$. 
The drag coefficients in all directions can be measured by tracking the position and orientation of rod-like proteins and applying the Fluctuation-Dissipation theorem. The measured drag coefficients  can then be used to compute the inter-leaflet friction and the membrane viscosity. 
\backsection[Supplementary data]{\label{SupMat}
\hypertarget{Supp}{Supplementary materials} are available at ...}


\backsection[Funding]{This study is supported by National Science Foundation under Career Grant No. CBET-1944156.}

\backsection[Declaration of interests]{The authors report no conflict of interest.}

%
\backsection[Author ORCIDs]{
\par
\noindent
\orcidlink{0000-0002-1097-8497} Wenzheng Shi \url{orcid.org/0000-0002-1097-8497}
\par
\noindent
\orcidlink{0000-0001-9247-9351} Moslem Moradi \url{orcid.org/0000-0001-9247-9351}
\par
\noindent
\orcidlink{0000-0002-3811-0067} Ehssan Nazockdast \url{orcid.org/0000-0002-3811-0067}
}
\appendix
\section{}\label{sec:appA}

Here, we outline the fundamental solutions to the system of  \cref{eq:Eqs} in response to a point-force, $\mathbf{f}^\text{ext}$, on the membrane at position $(\theta_0,\phi_0)$ on the outer layer of the sphere, where $\theta \in [0,\pi]$ is the polar angle and $\phi \in [0,2\pi)$ is the azimuthal angle, defined in \cref{fig:illustration}. The equations after including the externally applied point force are:
\begin{subequations}
\begin{align}
&\eta^{\pm}\nabla^{2}\mathbf{u}^{\pm}-\nabla p^{\pm}=\mathbf{0}, \\
&\nabla\cdot\mathbf{u}^{\pm}=0, \\
&\eta_{m}\left(\Delta_\gamma \mathbf{u}_{m}^{\text{out}}
+K\mathbf{u}_{m}^{\text{out}}-\frac{1}{b^{2}}(\mathbf{u}_{m}^{\text{out}}-\mathbf{u}_{m}^{\text{in}})\right)-\nabla_\gamma p_{m}^{\text{out}} +\mathbf{T}^{\text{out}}|_{r=R}+\mathbf{f}^{\text{ext}}{\delta} (\theta_0,\phi_{0})=\mathbf{0}, \\
&\nabla_\gamma\cdot \mathbf{u}_{m}^{\text{out}}=0, \\ 
&\eta_{m}\left(\Delta_\gamma \mathbf{u}_{m}^{\text{in}}
+K\mathbf{u}_{m}^{\text{in}}-\frac{1}{b^{2}}(\mathbf{u}_{m}^{\text{in}}-\mathbf{u}_{m}^{\text{out}})\right)-\nabla_\gamma p_{m}^{\text{in}} +\mathbf{T}^{\text{in}}|_{r=R}=\mathbf{0}, \\
&\nabla_\gamma\cdot \mathbf{u}_{m}^{\text{in}}=0,
\end{align}
\label{eq:AppEqs}
\end{subequations}
where $\delta(\theta_0,\phi_0)$ is Dirac delta function. 
The analytical solution to suspended monolayer spherical membrane was provided by \citet{Henle2010}. Here we extend their results to solve for the \cref{eq:AppEqs}. The velocity field at an arbitrary point $(\theta,\phi)$ in the outer-layer is 
$\mathbf{u}(\theta,\phi)=\mathbf{G}(\theta,\phi,\theta_0,\phi_0)\cdot \mathbf{f} (\theta_0,\phi_0)$. Writing this expression in matrix form gives:
\[
\begin{bmatrix}
    {u}_{\theta} \\
    {u}_{\phi} \\
\end{bmatrix}
=
\frac{1}{4\pi\eta_{m}}
\begin{bmatrix}
G_{\theta\theta} & G_{\theta\phi} \\
G_{\phi\theta} & G_{\phi\phi} \\
\end{bmatrix}
\cdot
\begin{bmatrix}
    {f}_{\theta}^\text{ext}(\theta_0,\phi_0) \\
    {f}_{\phi}^\text{ext} (\theta_0,\phi_0) \\
\end{bmatrix},
\]
where 
\begin{subequations} \label{eq:green}
\begin{align}
     G_{\theta\theta}=\sum_{l=2}^{\infty} &\frac{2l+1}{s_{l}l(l+1)}
    \Big{(} 
    -P^{2}_{l}(\cos{\psi})\sin^{-2}{\psi}\sin{\theta}\sin{\theta_{0}}\sin^{2}(\phi-\phi_{0}) \\
    \nonumber
    &-P^{1}_{l}(\cos{\psi})\sin^{-1}{\psi}\cos{(\phi-\phi_{0})}
    \Big{)}\\
    G_{\theta\phi} = \sum_{l=2}^{\infty}& \frac{2l+1}{s_{l}l(l+1)}
    \Big{(} P^{2}_{l}(\cos{\psi})\sin^{-2}{\psi}
    (-\cos{\theta}\sin{\theta_{0}}+\sin{\theta}\cos{\theta_{0}}\cos{(\phi-\phi_{0})}) \\
    \nonumber 
    &\cdot
    \sin{\theta_{0}}\sin{(\phi-\phi_{0})} -P^{1}_{l}(\cos{\psi})\sin^{-1}{\psi}\cos{\theta_{0}}\sin{(\phi-\phi_{0})}
    \Big{)}\\
    G_{\phi\theta} =\sum_{l=2}^{\infty}& \frac{2l+1}{s_{l}l(l+1)}
    \Big{(}
    P^{2}_{l}(\cos{\psi})\sin^{-2}{\psi}
    (\sin{\theta}\cos{\theta_{0}}-\cos{\theta}\sin{\theta_{0}}\cos{(\phi-\phi_{0})}) \\
    \nonumber
    &\sin{\theta}\sin{(\phi-\phi_{0})}
    +P^{1}_{l}(\cos{\psi})\sin^{-1}{\psi}\cos{\theta}\sin{(\phi-\phi_{0})}
    \Big{)},\\
    G_{\phi\phi} = \sum_{l=2}^{\infty}& \frac{2l+1}{s_{l}l(l+1)}
    \Big{(}
    P^{2}_{l}(\cos{\psi})\sin^{-2}{\psi}
    (-\cos{\theta}\sin{\theta_{0}}+\sin{\theta}\cos{\theta_{0}\cos{(\phi-\phi_{0})}}) \\
    \nonumber 
    &\cdot(-\sin{\theta}\cos{\theta_{0}}+\cos{\theta}\sin{\theta_{0}\cos{(\phi-\phi_{0})}}) \\
    \nonumber 
    &-P^{1}_{l}(\cos{\psi})\sin^{-1}{\psi}
    (\sin{\theta}\sin{\theta_{0}}+\cos{\theta}\cos{\theta_{0}\cos{(\phi-\phi_{0})}})
    \Big{)},
    \end{align}
\end{subequations}
\begin{equation}
\cos{\psi}=\cos{\theta}\cos{\theta_{0}}+\sin{\theta}\sin{\theta_{0}}\cos{(\phi-\phi_{0})},
\end{equation}
and
\begin{equation} \label{eq:sl}
s_{l}=l^{2}+l-2+\frac{R}{\ell^{+}}(l+2)+
\left(\left(
l^{2}+l-2+
\frac{R}{\ell^{-}}
\left(
\frac{(l-1)+(l+2)\left(1-\frac{H}{R}\right)^{2l+1}}{1-\left(1-\frac{H}{R}\right)^{2l+1}}
\right)
\right)^{-1}
+\left(\frac{R^{2}}{b^{2}}\right)^{-1}
\right)^{-1}.
\end{equation}
Here, $P_{l}^{m}(\cos{\psi})$ is the Associated Legendre polynomials with degree $l$ and order $m$, $\ell^{\pm}=\eta_{m}/\eta^{\pm}$, $b=\sqrt{\eta_{m}/\mu}$, where $\mu$ is the inter-leaflet drag coefficient, $H$ is the depth of the inner fluid, and $R$ is the radius of the sphere. Note that the summation of $l$ in Green's function starts from $l=2$ where we exclude the rigid-body rotation term $l=1$ because we only consider \emph{relative} motion of filament with respect to the spherical membrane \citep{Henle2010, samanta2021vortex, shi2022hydrodynamics}. 
\par
When the interior fluid is very thin, $H/R\ll1$, \cref{eq:sl}, to the first order approximation, can be simplifies to:
\begin{equation} \label{eq:sl2}
    s_{l}\approx l^{2}+l-2+\frac{R}{\ell^{+}}(l+2)+\frac{R^{2}}{\ell^{-}H+b^2}.
\end{equation}
We recover the same expression for $s_l$ if we set $H=0$, and with it  $\mathbf{u}_m^\text{in}=\mathbf{0}$, and substituting $b$ with $b^\star=\sqrt{\ell^{-}H+b^2}$. 

\bibliographystyle{jfm}
\bibliography{jfm}

\end{document}